\def\ttitle{Atom-Chip Compatible Optical Lattice}
\def\kkeywords{
  optical lattice,
  atom-chip
}

\documentclass[9pt,twoside,twocolumn]{opticajnl}
\journal{opticajournal} 
\setboolean{shortarticle}{true}

\newboolean{distrostatement}
\setboolean{distrostatement}{true}

\usepackage{ifthen}
\usepackage{charter}
\usepackage{cancel}
\usepackage{makeidx}
\usepackage{amsthm}
\usepackage{amssymb}
\usepackage{amsfonts}
\usepackage{graphicx}
\usepackage{svg}
\usepackage{MnSymbol}
\usepackage{float}
\usepackage[numbers,sort&compress]{natbib}
\usepackage{tikz}
\usepackage{IEEEtrantools}

\usepackage{array}
\usepackage{amsmath}
\usepackage{tabularx}
\usepackage{yfonts}
\usepackage{fancyhdr}
\usepackage{authblk}
\usepackage[nolist]{acronym}
\usepackage{./sty/shortcuts} 
\usepackage{eso-pic}
\usepackage{color}
\usepackage{type1cm}
\usepackage{import}      
\usepackage{mathtools, nccmath}
\usepackage{gensymb}
\usepackage{soul}

\DeclareMathOperator{\micro}{\mu}

\hypersetup{
    naturalnames=true,
    colorlinks=true,
    linkcolor=blue,
    pdfpagemode=UseNone,
    pdfstartview=FitH,
    pdftitle={\ttitle},
    pdfauthor=Air Force Research Laboratory,
    pdfsubject={Optical Lattice, Atom-Chip},
    pdfkeywords={\kkeywords},
}

\newcolumntype{L}[1]{>{\raggedright\let\newline\\\arraybackslash}m{#1}}
\newcolumntype{C}[1]{>{\centering\let\newline\\\arraybackslash}m{#1}}
\newcolumntype{R}[1]{>{\raggedleft\let\newline\\\arraybackslash}m{#1}}

\newcommand{\distA}[1]{%
  Approved for public release; distribution is unlimited.  Public Affairs %
  release approval %
  #1.
}

\ifthenelse{\boolean{distrostatement}}
{
\pagestyle{fancy}
\fancyfoot{}
\fancyfoot[R]{\thepage}
\fancyfoot[L]{
  \footnotesize
  \centering{\distA{AFRL20250050}}
}
}{}

  {\color{red}}%
  {}

\begin{acronym}
  \acro{ACL}{atom-chip lattice}
  \acro{SCL}{simple cubic lattice}
  \acro{DRSC}{degenerate Raman sideband cooling}
  \acro{PGC}{polarization gradient cooling}
  \acro{TOF}{time-of-flight}
  \acro{MOT}{magneto-optical trap}
\end{acronym}

\title{\ttitle}

\author[1]{Robert Leonard}
\author[2]{Spencer E. Olson}
\affil[1]{\footnotesize Space Dynamics Laboratory, Quantum Sensing \& Timing, North Logan, UT 84341, USA}
\affil[2]{Air Force Research Laboratory, Kirtland AFB, NM 87117, USA}

\affil[*]{qst@afrl.af.mil}

\begin{abstract}
A lattice beam configuration which results in an isotropic 3D trap near the
surface of an atom chip is described.  The lattice is formed near the
surface of a reflectively coated atom chip, where three incident beams and three
reflected beams intersect.  The coherent interference of these six beams form a
phase-stable optical lattice which extends to the surface of the atom chip.  The
lattice is experimentally realized and the trap frequency is measured.
Degenerate Raman sideband cooling is performed in the optical lattice, cooling
80 million atoms to $1.1\uK$.
\end{abstract}

\setboolean{displaycopyright}{false} 

\begin{document}

\setboolean{showcomment}{true}

\maketitle


\section{Introduction}\label{sec:intro}
Magnetic and optical traps are powerful tools in atomic physics and these
two trap types have unique strengths and weaknesses.  Magnetic traps
offer greater flexibility to engineer potentials tailored to specific
applications.  Moreover, magnetic traps formed by an atom chip are well suited
for use in portable and commercial devices due to their compact size, localized
fields, low power consumption, and mature fabrication techniques~\cite{Keil2016}.
However magnetic traps only trap atoms in a low-field seeking Zeeman state,
whereas optical traps are insensitive to the Zeeman state.  Moreover, optical
lattices commonly trap atoms in the Lamb-Dicke regime, as is required when
performing sideband cooling.

An optical lattice which spatially overlaps with the trap formed by an atom
chip could enable capabilities such as the rapid production of a cold atom gas in
a magnetic trap, and even production of Bose-Einstein condensates without the
use of evaporative cooling\cite{Olshanii2002}.
By alternating between sideband cooling in an optical lattice and
magnetic compression and reshaping, high lattice site occupation and
adiabatic transfer into a magnetic trap could be achieved.  This would enable
lossless loading of atoms into low-energy states of magnetic traps as well as
rapid Bose-Einstein condensate production similar to a previous demonstration in a cross-dipole
trap~\cite{Urvoy2019}.

Spatial overlap of an optical lattice with the trap created by an atom chip,
would typically require an optical lattice to be formed within a few hundred
micrometers of the atom chip.  Two groups have reported trapping atoms in an
optical lattice near an atom chip: Gallego \textit{et al.}~\cite{Gallego:09}
were the first to combine an atom chip with a 1D optical lattice, while
Straatsma \textit{et al.}~\cite{Straatsma:15} extended this to a 3D optical
lattice.

The lattice used by Straatsma \textit{et al.} consisted of three, mutually
orthogonal, retro-reflected beams.  The light used to form the lattice had
a short coherence length, such that each beam coherently interfered only with its
retro-reflected counterpart.  We will refer to this lattice design as the
\ac{SCL}.  With the \ac{SCL}, two lattice beams must travel parallel to the atom
chip.  To achieve overlap with an atom chip magnetic trap, the chip-beam
separation was $100 \um$. To minimize diffraction, incident lattice beams were
displaced from the chip surface using periscoping optics bonded to the chip.
The lattice beams used by Straatsma \textit{et al.} were focused to a waist of
$100 \um$.

In this work we present an alternative beam configuration which forms a 3D optical lattice
near the surface of an atom chip.  The lattice is formed by reflecting three
incident beams off the surface of a reflectively coated atom chip.  This design
allows the lattice to be easily integrated into mirror \ac{MOT}
systems~\cite{Grossmann1998}, commonly used to load atom chips.
Because the lattice uses beams that are reflected from the atom chip, the
design is not constrained by diffraction from the atom chip.  Consequently, the
lattice extends to within a wavelength of the atom chip. Furthermore, the
lattice avoids the use of on-chip periscoping optics, thereby simplifying
fabrication and improving optical access.  Large beams may be used to create the
lattice, which simplifies alignment and increases capture volume.  Because the
lattice forms at the surface of the atom chip, beam scatter from the chip
may be imaged to further assist with beam alignment.    We refer to this lattice
as the \ac{ACL}.

In Sec.~\ref{sec:theory}, we describe the beam geometry used to create the
\ac{ACL}. We present an expression for the lattice potential as well as formulas
for the lattice properties.  We compare the properties of the \ac{ACL} to the \ac{SCL}.
We conclude this section with a brief discussion on
the phase stability of the lattice.  In Sec.~\ref{sec:experiment}, we describe an experimental
implementation of the lattice.  We present measurements of the lattice frequency,
and demonstrate \ac{DRSC}.

\section{Atom Chip Lattice Theory}\label{sec:theory}
The far-off-resonant light that makes up an optical lattice interacts with the light induced
dipole moment of an atom giving rise to the ac-Stark shift.  The ac-Stark shift
is given by
\begin{equation}
  \label{eq:stark}
  U\left(\vec{r}\right) = \frac{-1}{2} \, \alpha \left(\omega_L\right) \, \langle \vec{E}\left(\vec{r}\right) \, \vec{E}^{*}\left(\vec{r}\right) \rangle
\end{equation}
where the angled brackets denote a time-average, $\alpha ( \omega_L)$
is the dynamic polarizability created through the interaction between the atom
and light with angular frequency $\omega_L$, $\vec{E}$ is the electric field of
the light, and $\vec{E}^{*}$ denotes its complex-conjugate.

\begin{figure}[htb!]
  \centering
  \def\svgwidth{\columnwidth}
\begingroup%
  \makeatletter%
  \providecommand\color[2][]{%
    \errmessage{(Inkscape) Color is used for the text in Inkscape, but the package 'color.sty' is not loaded}%
    \renewcommand\color[2][]{}%
  }%
  \providecommand\transparent[1]{%
    \errmessage{(Inkscape) Transparency is used (non-zero) for the text in Inkscape, but the package 'transparent.sty' is not loaded}%
    \renewcommand\transparent[1]{}%
  }%
  \providecommand\rotatebox[2]{#2}%
  \newcommand*\fsize{\dimexpr\f@size pt\relax}%
  \newcommand*\lineheight[1]{\fontsize{\fsize}{#1\fsize}\selectfont}%
  \ifx\svgwidth\undefined%
    \setlength{\unitlength}{285.12375718bp}%
    \ifx\svgscale\undefined%
      \relax%
    \else%
      \setlength{\unitlength}{\unitlength * \real{\svgscale}}%
    \fi%
  \else%
    \setlength{\unitlength}{\svgwidth}%
  \fi%
  \global\let\svgwidth\undefined%
  \global\let\svgscale\undefined%
  \makeatother%
  \begin{picture}(1,0.64285344)%
    \lineheight{1}%
    \setlength\tabcolsep{0pt}%
    \put(0.89665124,0.11551307){\color[rgb]{0,0,0}\makebox(0,0)[t]{\lineheight{0}\smash{\begin{tabular}[t]{c}\textbf{3-D MOT}\end{tabular}}}}%
    \put(0.1146922,0.09499077){\color[rgb]{0,0,0}\makebox(0,0)[t]{\lineheight{0}\smash{\begin{tabular}[t]{c}\textbf{2-D MOT}\end{tabular}}}}%
    \put(0,0){\includegraphics[width=\unitlength,page=1]{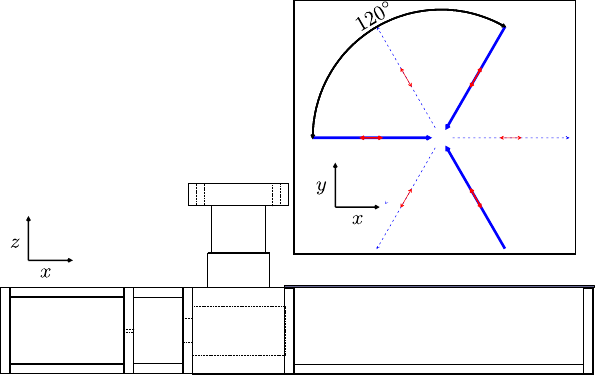}}%
    \put(0.81592403,0.50625476){\color[rgb]{0,0,0}\makebox(0,0)[lt]{\lineheight{1.10000002}\smash{\begin{tabular}[t]{l}Top view\end{tabular}}}}%
    \put(0,0){\includegraphics[width=\unitlength,page=2]{lattice-layout_svg-tex.pdf}}%
  \end{picture}%
\endgroup%

  \caption[Pointing and polarization vectors used for the \ac{ACL}.]{
    Poynting and polarization vectors used for the \ac{ACL}.
  }
  \label{fig:pointing_side}
\end{figure}

The \ac{ACL} consists of three, coherent, red-detuned beams with equal intensity.
All beams are directed towards a reflectively coated atom chip at an angle of
incidence of $\phi = \tan^{-1}\sqrt{2}$.  The beams are
aligned to intersect at the surface of the atom chip.  We define the z-axis as
normal to the atom chip.  Incident beams are separated by $120\degree$ in the
xy-plane.  In this configuration, incident beams are mutually orthogonal.  All
beams are linearly polarized with P-polarization.  Fig.~\ref{fig:pointing_side}
illustrates the Poynting and polarization vectors for the beams used to create
the \ac{ACL}.

%
%
The potential created by this lattice is
\vspace{-.25em}
\begin{multline}
  \label{eq:three_beam_potential}
  U(\vec{r}) = \frac{-1}{2 \, c \, \epsilon_0} \, I_0 \, \alpha \left( \omega_L \right)
    \times\Bigg\{ -\frac{4}{3} \, \cos^2\left( k \, z \, \cos \phi\right) \bigg[\\
    -3 + 2 \cos\left(\frac{k \, x}{\sin \phi}\right) \cos\left(\frac{k \, y}{\tan \phi}\right)
      + \cos \left( k \, y \, \tan \phi \right) \bigg]\\
    + \frac{8}{3} \sin^2\left( k z \cos \phi \right) \bigg[ 3 + 4 \cos \left( \frac{k \, x}{\sin \phi} \right)
      \cos\left( \frac{k \, y}{\tan \phi} \right)\\
      + 2 \cos\left(k \, y \, \tan \phi \right) \bigg] \Bigg\}
\end{multline}
where $\phi = \tan^{-1}\sqrt{2}$ and $I_0$ is the intensity of the
beams used to create the lattice.

\begin{figure}[htb!]
  \includegraphics[clip, trim=0cm 0.1cm 0cm 0.1cm, width=\columnwidth]{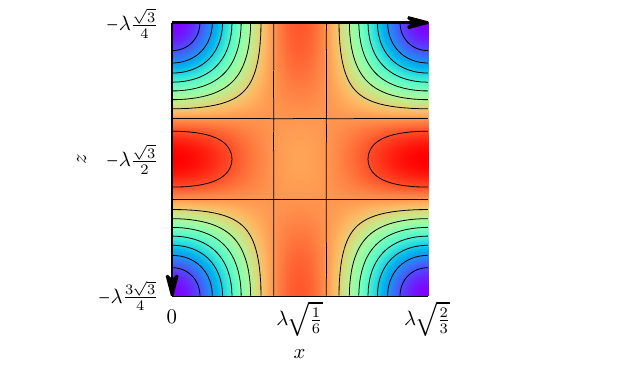}
  \includegraphics[clip, trim=0cm 0.05cm 0cm 0.1cm, width=\columnwidth]{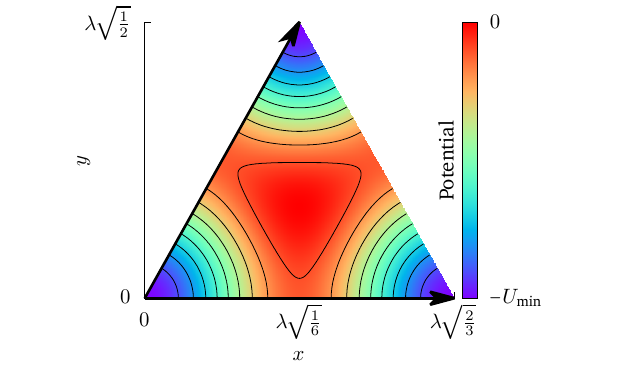}
  \caption[Lattice potential in $\vec{r}_1 \times \vec{r}_2$ and
           $\vec{r}_1 \times \vec{r}_3$.]{
    Lattice potential in the space spanned by principle vectors
    $\vec{r}_1 \times \vec{r}_3$ at $y = 0$ (top),  and
    $\vec{r}_1 \times \vec{r}_2$ at $z = - \lambda \sqrt{3}/4$ (bottom).  The
   spaces shown are chosen to include the potential minima of the lattice.
  }
  \label{fig:xy-plane}
\end{figure}

The resulting lattice has isotropic trap frequencies.  There is translational
symmetry along the z-axis with spatial periodicity of $\lambda \sqrt{3}/2$.
In the xy-plane, local minima form equilateral triangles arranged in a hexagonal
pattern.  Minima are separated by $\lambda \sqrt{2/3}$ in the xy-plane.  The
potential formed by the \ac{ACL} is illustrated in Fig.~\ref{fig:xy-plane} while
the properties of the lattice are summarized in Tab.~\ref{table:properties}. The
barrier described in Tab.~\ref{table:properties} refers to the minimum potential
barrier between adjacent lattice sites.

\begin{table}
\centering
{\renewcommand{\arraystretch}{2}
\begin{tabular}{|C{1.2cm}|C{2.5cm}|c|}
\hline
                          & \ac{SCL}                                                          & \ac{ACL}                                                                  \\ \hline
$\ssc{U}{min}$            & $-6 \frac{\alpha(\omega_L) \, I_0}{c \, \epsilon_0}$                         & $-12 \frac{\alpha(\omega_L) \, I_0}{c \, \epsilon_0}$                                \\ \hline
$f$                       & $\frac{1}{\lambda} \sqrt{\frac{-2 \, \ssc{U}{min}}{3 \, m}}$               & $\frac{1}{\lambda} \sqrt{\frac{-2 \, \ssc{U}{min}}{3 \, m}}$                       \\ \hline
$\frac{\ud^4 U}{\ud x^4}$ & $\frac{8}{3} \ssc{U}{min} k^4$                                             & $\ssc{U}{min} k^4$                                                                 \\ \hline
$\frac{\ud^4 U}{\ud z^4}$ & $\frac{8}{3} \ssc{U}{min} k^4$                                             & $\frac{8}{9} \ssc{U}{min} k^4$                                                     \\ \hline
Barrier                   & $\frac{1}{3} \ssc{U}{min}$                                                 & $\frac{4}{5} \ssc{U}{min}$                                                         \\ \hline
                          & $\vec{r}_1 \, = \, \lambda \left \langle \frac{1}{2}, 0, 0 \right \rangle$ & $\vec{r}_1 \, = \, \lambda \left \langle \sqrt{\frac{2}{3}}, 0, 0 \right \rangle$  \\
Principle Vectors         & $\vec{r}_2 \, = \, \lambda \left \langle 0, \frac{1}{2}, 0 \right \rangle$ & $\vec{r}_2 \, = \, \lambda \left \langle \sqrt{\frac{1}{6}}, \sqrt{\frac{1}{2}}, 0 \right \rangle$   \\
                          & $\vec{r}_3 \, = \, \lambda \left \langle 0, 0, \frac{1}{2} \right \rangle$ & $\vec{r}_3 \, = \, \lambda \left \langle 0, 0, \frac{\sqrt{3}}{2} \right \rangle$  \\ \hline
\end{tabular}
}
\caption[Lattice Properties]{
  Properties of the \ac{SCL} and the \ac{ACL}.
}
\label{table:properties}
\vspace{-1.5em}
\end{table}

\Ac{ACL} properties are independent of the phases of the incident
beams as well as displacements in the reflecting mirror.  A well-established
rule\cite{Grynberg93} for lattice-phase stability states that a lattice using
coherently interfering beams to confine atoms in $n$ dimensions must have no
more than $n+1$ independent phases.  The \ac{ACL} may appear to violate this rule,
as the lattice is formed by the coherent interference of six beams.  However, the
\ac{ACL} contains only four independent phases: three phases from
the incident beams, with a fourth phase from the reflection from the surface.
Because all three incident beams reflect off the same point of the same
surface, all three beams receive a common phase shift upon reflection.
Relative phase shifts in the incident beams result in a translation of the
lattice in the xy-plane.  Movement of the mirror results in a translation
along the z-axis.

While the \ac{ACL} is phase stable, when the angles of incidence of the lattice
beams differ, the lattice loses spatial uniformity.  Properties of the
lattice, such as depth and frequency, vary along the axis normal to the
reflective surface, while the properties of the lattice are constant in the
plane parallel to the reflective surface.  The spatial frequency of the property
variations increases with increasing difference in the angles of incidence.
Crucially, however, the magnitude of these variations are independent of the
difference in the angles of incidence.  Consequently, the maximum and minimum
trap depth and frequency do not depend strongly on beam alignment. This enables
optimized sideband cooling which is robust to beam alignment stability.

\section{Experimental Realization}\label{sec:experiment}

Experimental realization of the lattice was achieved using $^{87}$Rb. The system
used for this experiment (Fig.~\ref{fig:pointing_side}) is similar to the one described in Ref.~\cite{Squires16}.
Atoms are prepared in a mirror \ac{MOT} with the cloud center $4.18 \mm$ below
the atom chip.  Before loading into the optical lattice, the magnetic fields are
ramped, moving the cloud $944 \um$ below the atom chip.  The atoms are further
cooled to $8.4 \uK$ through \ac{PGC}.

The lattice is formed using three beams from a 1:3 fiber splitter.
Polarization-maintaining
fibers are used to ensure the lattice beams maintain P-polarization.
The beams are launched from fiber collimators mounted to custom pedestals angled
to launch the beams with an angle of $35.3\degree$ above the horizontal.  The beams have $4\mm$
beam waist and $39 \mW$, $45 \mW$, and $44 \mW$ of power. The lattice is
tuned $13.25 \GHz$ red of the
$| 5\mathrm{S}_{1/2}, \mathrm{F}=2\rangle \rightarrow | 5\mathrm{P}_{3/2}, \mathrm{F}=2\rangle$
resonance.  This is similar to detunings seen in other sideband cooling
experiments~\cite{AJKerman,Wei2017}, where the small detunings allow the lattice
to drive the Raman transitions between degenerate states.  The heating rate of
atoms held in the lattice was found to be $430 \nK/\ms$.

Alignment of the lattice is achieved with the help of an absorption imaging
system with imaging axis oriented normal to the atom chip.  The imaging beam is
retro-reflected from the atom chip and directed into the imaging system via a
beam splitting cube.  The atoms are imaged
following \ac{PGC}.  Lattice beams are aligned so that the center of the beams
impinge on the atom chip at the location of the atoms following \ac{PGC}.  This
is easily achieved through imaging the light scattered by the lattice beams off
the surface of the atom chip.

\begin{figure}[htb!]
  \hspace{-0.5cm}
  \includegraphics[height=.18\textwidth]{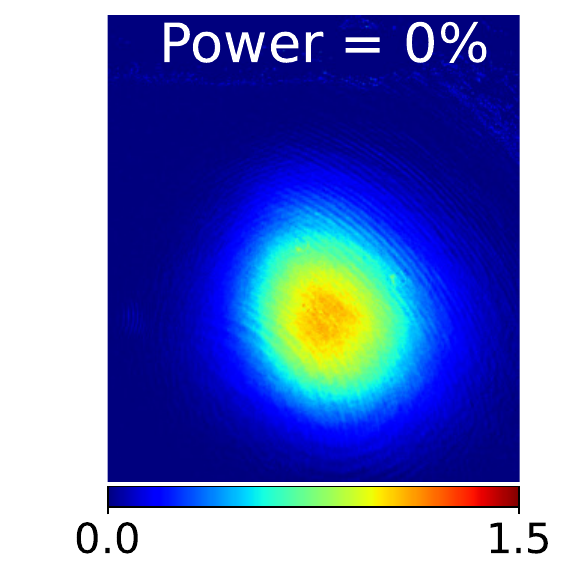}
  \hspace{-0.2cm}
  \includegraphics[height=.18\textwidth]{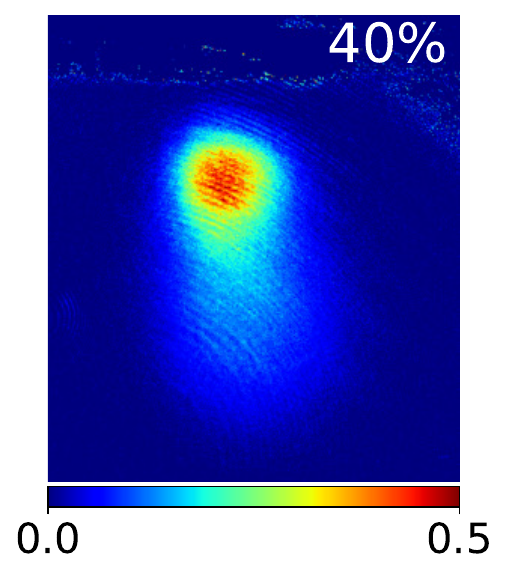}
  \hspace{-0.1cm}
  \includegraphics[height=.18\textwidth]{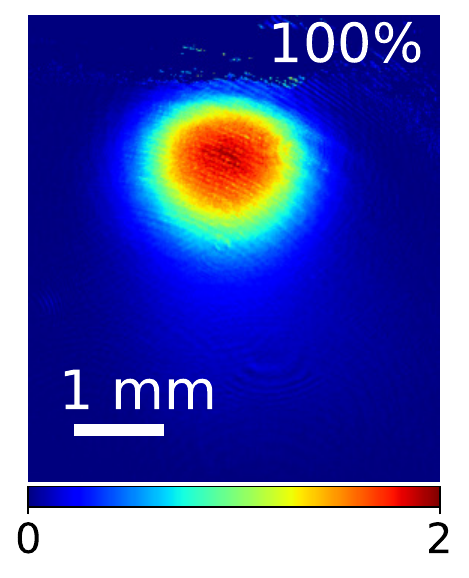}
  \caption[Absorption image of atoms in lattice.]{
    Absorption image of atoms after a being held for $15\ms$ in
    the optical lattice with the indicated fractional optical power, followed by a
    $10\ms$ \ac{TOF}.  The fraction of atoms
    transferred from the \ac{MOT} into the optical lattice was calculated by
    fitting the optical density to two Gaussians centered at the expected locations
    of the trapped and untrapped atom clouds.  From this analysis we calculate
    that $54\%$ of the atoms are transferred from the \ac{MOT} into the lattice
    when the lattice beams are limited to $40\%$ power, while $>90\%$ of the
    atoms are transferred at full power.}
  \label{fig:transfered}
\end{figure}

We measure the fraction of atoms transferred from the \ac{MOT} into the lattice
using a horizontally oriented imaging system.  Atoms are held in the lattice
for $15\ms$, which is sufficient for the untrapped atom cloud to spatially separate
from the trapped atoms while preventing atoms from leaving the imaging field of
view. The atoms are then released from the lattice for a short \ac{TOF} and
imaged.  The resulting optical density, shown in Fig.~\ref{fig:transfered}, is
fit to two Gaussians.  From this analysis we calculate the fraction of atoms
captured in the lattice.  Using this technique we find that we transfer $>90\%$
of the atoms from the \ac{MOT} into the optical lattice.

The lattice trap frequency was measured using parametric heating.  Heating was
achieved by modulating the intensity of the lattice beams, while temperature was
measured through \ac{TOF}.  Parametric heating measurements
show a trap frequency of $49.7\kHz$, close to the expected
frequency of $46.6\kHz$.

The motional energy of atoms in the \ac{ACL} is quantized such that the
motional energy quanta are greater than the recoil energy
transferred by near-resonant photon scatter; this regime is known as the
Lamb-Dicke regime.  Atoms in this limit are characterized by the Lamb-Dicke
parameter, $\eta = \sqrt{\ssc{E}{R}/\hbar \, \ssc{\omega}{trap}}$; for the work
presented here, $\eta=0.27$.  In the Lamb-Dicke regime,
the probability that resonant photon scatter changes the motional energy of the
atoms is reduced.  This property of the Lamb-Dicke regime is exploited during
the optical pumping stage of \ac{DRSC}.

In \ac{DRSC}, a small magnetic field is applied such that the Zeeman
splitting between $m_F$ states equals the motional energy quantization
created by the optical lattice.  Off-resonant light drives the 2-photon
degenerate Raman transitions between the $|v=n, m_F = F\rangle$ and
$|v=n-1, m_F = F-1\rangle$ states, where $v$ is the quantum number describing the
motional state of the trapped atom.  This effectively enables the atoms to trade
Zeeman energy for motional energy.  Because this process is coherent, heating
and cooling both occur with equal probability.  Optical pumping is used to
create a ``ratchet,'' allowing the atoms to repeatedly trade higher Zeeman
energy for lower motional energy, while suppressing the reverse process.

\begin{figure}[htb!]
  \centering
    \includegraphics[clip, trim=0cm 0.1cm 0cm 0.52cm,width=1\columnwidth]{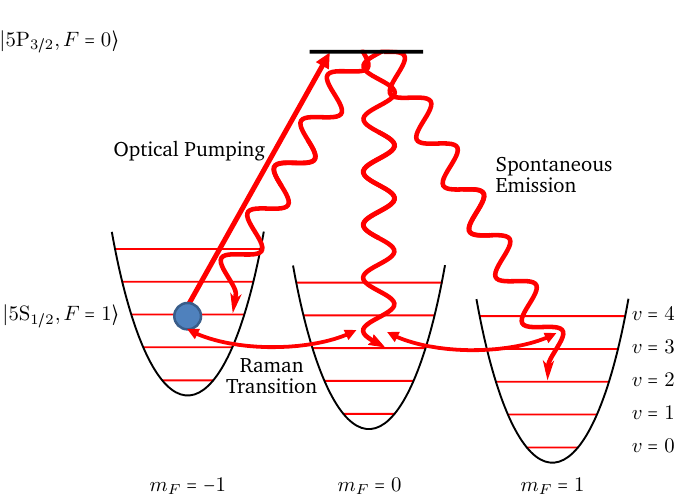}
  \caption[This is how \ac{DRSC} works.]{
    A precisely tuned Zeeman shift creates a degeneracy between $|v=n, m_F = F\rangle$ and
$|v=n-1, m_F = F-1\rangle$ states.  A two-photon Raman transition drives atoms
between these degenerate states, trading Zeeman energy for motional energy.
Optical pumping creates a ``ratchet,'' allowing the atoms to repeatedly trade
higher Zeeman energy for lower motional energy, while suppressing the reverse
process.
  }
  \label{fig:drsc}
\end{figure}

In this experiment, our atoms are loaded into the
$|5\mathrm{S}_{1/2}, \mathrm{F}=1\rangle$ ground state.  As depicted in
Fig.~\ref{fig:drsc}, optical pumping light
is tuned to the  $|5\mathrm{S}_{1/2}, F=1\rangle \rightarrow | 5\mathrm{P}_{3/2}, \mathrm{F}=0\rangle$
resonance, while polarization and magnetic field direction are chosen so that the
optical pumping photons are $\ssc{\sigma}{+}$ polarized with respect to the atoms.
Because $|5\mathrm{P}_{3/2}, \mathrm{F}=0\rangle$ has only a single Zeeman state,
$m_F=0$, optical pumping photons only scatter when atoms are the highest Zeeman
energy, lowest motional energy, state: $|5\mathrm{S}_{1/2}, \mathrm{F}=1, m_F=-1\rangle$.
After scattering into $|5\mathrm{P}_{3/2}, \mathrm{F}=0\rangle$ these atoms decay
back into the $|5\mathrm{S}_{1/2}, \mathrm{F}=1\rangle$ ground state where
further cooling may occur.  Because atoms are trapped in the Lamb-Dicke regime,
any reduction in motional energy tends to be preserved during the optical pumping.
This process is repeated until the atom is in both the lowest motional energy and
lowest Zeeman energy state, $|5\mathrm{S}_{1/2}, v=0, \mathrm{F}=1, m_F=1\rangle$.
Atoms in this state are dark to both optical pumping and Raman light.

In preparation for \ac{DRSC}, our atoms are cooled using \ac{PGC}.  Following
this, the atoms are pumped into the $|F=1\rangle$ state by tuning the
\ac{MOT} cooling light to the
$|5\mathrm{S}_{1/2}, \mathrm{F}=2\rangle \rightarrow |5\mathrm{P}_{3/2}, \mathrm{F}=2\rangle$
resonance, and tuning the \ac{MOT} repump light very far from resonance; the
atoms are pumped for $2\ms$.  \ac{MOT} and \ac{PGC}  light is
extinguished, while the intensity of the optical
lattice is ramped adiabatically over $300\us$ following the curve described
in Ref.~\cite{AJKerman}.  After loading into the lattice, the magnetic
field is ramped over $100\us$ to create the Zeeman-motional degeneracy.  Optical
pumping and repump light is applied for $14\ms$.  Following \ac{DRSC} the
optical lattice is extinguished and the magnetic field is zeroed.  The
temperature of the cloud is measured through \ac{TOF}.

\begin{figure}[htb!]
  \includegraphics[clip, trim=0cm 0.1cm 0cm 0.25cm,width=\columnwidth]{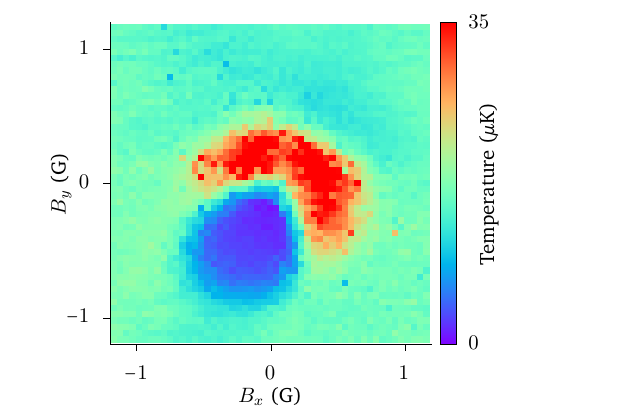}
  \caption[Magnetic field dependence of \ac{DRSC}]{
    Magnetic field dependence on temperature achieved during \ac{DRSC}.
    Temperature was measured for each bias field by measuring the cloud size
    following a $10\ms$ and $25\ms$ \ac{TOF}.  The field zero was determined
    through optimization of \ac{PGC}.
  }
  \label{fig:magnetic_dependence}
\end{figure}

The magnetic field dependence of the $\ac{DRSC}$ was explored by performing a
scan of the $x$ and $y$ magnetic bias fields.  For each bias field, the
temperature was calculated by measuring the cloud size following a $10\ms$ and
$25\ms$ \ac{TOF}.  This data, shown in Fig.~\ref{fig:magnetic_dependence},
shows clearly defined regions of Stokes heating and anti-Stokes cooling.  The
minimum temperature is seen at bias field of $140\mG$.

The best results were achieved by linearly ramping the intensity of the optical
lattice light, as well as the magnetic fields and optical pumping detuning
during the \ac{DRSC} stage.  The optimal experimental parameters were found
using a particle swarm optimizer modified to handle stochastic
data.  The \ac{DRSC} parameters were optimized to maximize phase-space density.
In the optimal configuration, the optical lattice intensity starts fully on and
ramps down to $71\%$ of max intensity.  The magnetic field strength starts at
$217\mG$ and ramps down to $56\mG$. Lastly, the optical pumping is initially
tuned $8.75\MHz$ blue of the
$|5\mathrm{S}_{1/2}, \mathrm{F}=1\rangle \rightarrow |5\mathrm{P}_{3/2}, \mathrm{F}=0\rangle$
resonance, and is increased to $14.5\MHz$ blue.  Following optimization, we
find that we cool 80 million atoms to $0.80\uK$ horizontally and $1.4\uK$
vertically.


As a preliminary test of the application of an \ac{ACL} for lossless loading
into atom-chip traps, as mentioned in Sec.~\ref{sec:intro},
we performed a sequence of \ac{DRSC} followed by optical pumping into
$\left|F=2, m_{F}=2\right>$ followed by \ac{DRSC}.  A single cycle returned the
atoms to $\sim 2\uK$.

\section{Conclusions}\label{sec:conclusion}
We have presented an isotropic 3D optical lattice design which extends to the
surface of an atom chip.  The lattice does not require on-chip optics and may be
formed using large beams, which enables increased capture volume while easing
alignment requirements.  Despite arising from the coherent interference of six
beams, the optical lattice is phase stable due to the common phase shift
applied to all three incident beams upon reflection.  The lattice presented here
represents a single example of a class of near-chip lattices which are formed
by the reflection of lattice beams off the surface of an atom chip.  Further
lattice designs including anisotropic 3D lattices, 2D lattices, and blue-detuned
lattices may be formed using this technique.

We experimentally demonstrated trapping of $^{87}$Rb in the \ac{ACL}.  We succeeded
in transferring more than $90\%$ of the atoms from the \ac{MOT} into this lattice.
We performed \ac{DRSC} in the atom-chip lattice, cooling 80 million
atoms to $1.1\uK$.  The ability to perform \ac{DRSC} in an optical lattice which
overlaps with the trap formed by an atom chip could enable high lattice site
occupation as well as adiabatic transfer into a magnetic trap through
alternating sideband cooling and magnetic trap compression stages.
Such a scheme could enable rapid production of Bose-Einstein condensates.

\begin{backmatter}
\bmsection{Funding}
This work was funded by the Air Force Office of Scientific Research under lab
task 22RVCOR017.

\bmsection{Disclosures} The authors declare no conflicts of interest.

\bmsection{Disclaimer}
The views expressed are those of the authors and do not necessarily reflect the
official policy or position of the Department of the Air Force, the Department
of the Defense, or the U.S. Government.
\end{backmatter}

\bibliography{lattice}
\end{document}